# On-column 2*p* bound state with topological charge ±1 excited by an atomic-size vortex beam in an aberration-corrected scanning transmission electron microscope

Huolin L. Xin[1*] and Haimei Zheng[1]

[1]Materials Sciences Division, Lawrence Berkeley National Lab, Berkeley 94720

[*]Correspondence should be addressed to hxin@lbl.gov

Running title: Vortex beams hosted in crystals …

## Abstract

Atomic-size vortex beams have great potential in probing materials' magnetic moment at atomic scales. However, the limited depth of field of vortex beams constrains the probing depth in which the helical phase front is preserved. On the other hand, electron channeling in crystals can counteract beam divergence and extend the vortex beam without disrupting its topological charge. Specifically, in this paper, we report atomic vortex beams with topological charge ±1 can be coupled to the 2*p* columnar bound states and propagate for more 50 nm without being dispersed and losing its helical phase front. We gave numerical solutions to the 2*p* columnar orbitals and tabulated the characteristic size of the 2*p* states of two typical elements, Co and Dy, for various incident beam energies and various atomic densities. The tabulated numbers allow estimates of the optimal convergence angle for maximal coupling to 2*p* columnar orbital. We also have developed analytic formulae for beam energy, convergence-angle, and hologram dependent scaling for various characteristic sizes. These length scales are useful for the design of pitch-fork apertures and operations of microscopes in the vortex-beam imaging mode.

# INTRODUCTION

Recently, the ability to create an electron beam that carries specified orbital angular momentum (Bliokh, et al., 2011; Herring, 2011; Idrobo & Pennycook, 2011; McMorran, et al., 2011a; McMorran, et al., 2011b; Schattschneider & Verbeeck; Uchida & Tonomura, 2010; Verbeeck, et al.; Verbeeck, et al., 2010) has generated immense interest for its potential to measure the magnetic moment of 3$d$ and rare earth elements at atomic scales (Schattschneider, 2008; Schattschneider, et al.; Xin & Muller, 2010). Mathematically, the complex wave function that contains an orbital angular momentum of $l\hbar$ can be written as $\Psi = e^{il\varphi} f(r)$, where $\varphi$ is the azimuthal angle, $r$ is the radial coordinate and $l$ is the topological charge. Vortex beams are defined as such beams with non-zero topological charges and with the wave amplitude going to zeros at $r = 0$ giving a phase singularity in the center (Allen, et al., 1992; Bazhenov, et al., 1990; Nye & Berry, 1974). Optical vortex beams have been studied intensively (Basistiy, et al., 1995; Berjersbergen, et al., 1994; Brand, 1999; He, et al., 1995; Heckenberg, et al., 1992; O'Neil, et al., 2002; Padgett, et al., 2004) and have found numerous applications in a wide range of fields such as optical tweezing (Curtis, et al., 2002; Grier, 2003) and quantum optics (Kapale & Dowling, 2005). In a similar manner to optical vortex beams, in an electron microscope, high-energy electrons that carry orbital angular momentums can be generated either using a helical phase plate (Uchida & Tonomura, 2010) or a computer-generated hologram (McMorran, et al., 2011b; Verbeeck, et al., 2010). The electron vortex beam has become more promising; as shown theoretically by Xin and Muller (Xin & Muller, 2010) and later by Schattschneider, Verbeeck and their coworkers experimentally (Schattschneider, et al.; Verbeeck, et al., 2011), sub-nanometer or even atomic-size beams with topological charge $\pm 1$ can be formed if the probe forming aperture at the pre-specimen focal plane is replaced by a pitchfork hologram aperture.

With the simplicity to create micron-scale features on metal foils using current lithography and focused ion beam technology, this concept can be realized in most modern instruments. Optical and electron vortex beams are in many ways similar both in their wave nature and their generation optics (Bliokh, et al., 2007). However, there are fundamental differences between them (Henderson, 1995). Apart from absorption, light and x-ray weakly interact with solids allowing the kinematic theory of light propagation to hold. Therefore, not much attention has been paid to study how multiple elastic scattering can modify the chirality of a vortex beam in the optical and x-ray community. Electrons however are strongly elastically scattered in materials. The strong elastic interaction with solids, also known as channeling in on-axis crystals (Hillyard, et al., 1993; Hillyard & Silcox, 1993; Kirkland, et al., 1987; Loane, et al., 1988), immediately raises alert that the desired chirality of the vortex beam may vanish or change even in very thin specimens. This, thus, requires scrutiny of vortex beam propagation in crystals. In this work we show that an electron vortex beam with a carefully chosen convergence angle can be efficiently couple to the on-column 2p state ($l$=1) that carries the same topological charge. The helical phase front of the incident vortex beam can be preserved while channeling on column. On the other hand, when propagating in vacuum, the helical phase is only preserved while the beam is within the depth of field.

The propagation of high-energy electrons in an ideal crystal is described by Bethe's Bloch-wave dynamical theory (Allen, et al., 2003; Bethe, 1928; Hirsch, et al., 1965). To solve the problem rigorously, all the reflections in the reciprocal lattice need to be considered. However, the Ewald sphere of high-energy electrons is relatively flat. Therefore, including only the reflections in the First Order Laue Zone in the calculation is a proper first order approximation (Geuens & Van

Dyck, 2002; Nellist & Pennycook, 1999; Pennycook & Jesson, 1990). Translating this into real space language, it means a crystal that makes of a matrix of atoms can be approximated by a 2-D array of rods. Each rod is described by an average attractive potential, as screened by the core and valance electrons. Each rod of screened charges can support cylindrical bound states on them in a similar manner to the atomic orbitals around atoms (Anstis, et al., 2003; Geuens & Van Dyck, 2002; Van Aert, et al., 2007). A tight-binding calculation can recover the dispersion relations. However if the overlap of states from neighbor columns are sufficiently small, as do the deep core orbitals of atoms in solids, the dispersion surface is nearly flat that columnar states can be considered to be isolated or well-defined (Anstis, et al., 2003; Pennycook & Jesson, 1990; Van Dyck & deBeeck, 1996). These well-defined columnar states carrying their specified orbital angular momentum (or topological charge) can be selectively exited by an incident beam that has the same angular momentum. For example, an ideal convergent beam when positioned on-column can only couple to the *s* states for its cylindrical symmetry. However, with the creation of on-column atomic-size vortex beams, columnar states with non-zero topological charges can be excited selectively. This selectivity allows the chirality of the incident vortex beam to be preserved on-column. As the 2*p* state has zero amplitude at $r=0$, it is a natural vortex state that is supported in a crystal. Therefore, it does not only preserve the topological charge but also preserves the vortex nature. The crystal columns serve as a depth-of-focus extender which counteracts the divergence nature of a convergent beam.

To provide a full picture of atomic-size vortex beam and its elastic interaction with a crystal, the rest of the paper is organized as follows. First, we will review how atomic-size vortex beam can be created in an electron microscopy and give the scalings of various characteristic dimensions of vortex beams as a function of the convergence angle. Second, we will review the scattering theory and laid down the radial Schrodinger equation needed for solving the columnar states. Third, we will show how the vortex beam can be coupled to the 2*p* state and channel on the column without losing its vortex nature.

# ATOMIC-SIZE VORTEX BEAM AND ITS FREE PROPAGATION

It has been shown that the Fourier transform of a pitchfork aperture can produce vortex beams with topological charge ±1 in the first side bands of the central beam (Allen, et al., 1992; Bazhenov, et al., 1990). In transmission electron microscope, (McMorran, et al., 2011b; Verbeeck, et al., 2010) experimentally verified this by diffracting off the aperture in the image plane. However, as shown by Idrobo and Pennycook, it is impractical to create sub-nanometer vortex beams using this approach (Idrobo & Pennycook, 2011). However, the Fourier transform can be carried out in reverse. By placing the pitchfork aperture in the back focal plane of the probe forming lens, atomic-size size vortex beams are formed at the front focal plane (Figure 1a)(Schattschneider, et al.; Xin & Muller, 2010). Figure 1b shows an example of the amplitude and phase of the central convergent beam and the first two side bands. As clearly indicated the phase map shown in Figure 1b, the two side beams have a helical phase front: the phase increases by $2\pi$ as it goes around the vortex core.

Figure 2 shows the peak radius ($R_{peak}$), the Rayleigh radius ($R_{Rayleigh}$) and the full width at half maximum (FWHM) of the vortex beam with topological charge one as a function of the convergence angle[1]. (Here following Rayleigh criterion, the Rayleigh radius is defined as the distance from the center to the first minimum of the side lobe.) These are various dimensions defined to characterize the size of the vortex beam. The same as the scaling for any convergent beam, the dimension of the beam is proportional to the wavelength of the incident electron and is inversely proportional to the convergent angle. The pre-factor differs in different measurements of the beam size. For example the pre-factor for the Rayleigh radius is 0.91. It is 50% larger than the more familiar factor—0.61—for a convergent beam form by a round aperture.

The phase around the vortex core is perfectly helical when the beam is at its optimal defocus. However, due to the divergence nature of a convergent beam the helical phase front can be destroyed. To illustrate this effect, Figure 3 shows the propagation of the wave front away from the perfect defocus plane. We find that, first the amplitude part of the vortex beam diverges in size as the beam propagates. Second, the phase part of the vortex beam becomes a spiral pattern. Third, the interferences from the central beam can strongly modify the side bands as the beam propagates further. At 20nm, we can see that the helical nature of the side beam is completely destroyed by the interferences coming from the divergent central beam. The depth at which such strong interference occurs can be estimated by $\frac{N}{2}\frac{\lambda}{\alpha_{max}^2}$, where the $\lambda$ is the wavelength of the incident electron, $\alpha_{max}$ is the convergence semi-angle and $N$ is the number of "white" streaks in the pitchfork aperture (For instance, $N = 8$ in Figure 1). (One would notice this is $N/4$ times the depth of focus ($\frac{2\lambda}{\alpha_{max}^2}$) of a convergent electron beam.) Therefore, given a chosen thickness and the convergence angle, the number of fringes needed in the aperture to ensure that the vortex side bands are not strongly influence by the central beam can be calculated by $N = d/(\frac{\lambda}{\alpha_{max}^2})$. ($d$ is the thickness. We assume the beam is focused at the center of the sample in depth). The results for three typical beam voltages are tabulated in Figure 4. It provides a guideline for designing pitchfork apertures.

---

[1] In the remaining text, convergence angle is the same as convergence semi-angle unless otherwise specified

# PROPAGATION IN CRYTALS

In last section, we have studied the propagation of vortex beam in vacuum. It is shown that the vortex beam diverges as it propagates away from the optimal focal plane. However, this phenomenon can be significantly modified by electron channeling. In this section, we will first develop the theory for solving the columnar bound states on atomic columns which are responsible for electron channeling. Then we will show that coupling to the $2p$ bound state helps prevent the vortex beam from diverging as it propagates on-column.

## The Scalar Relativistic-Corrected Schrodinger Equation

The speed of a 300 keV electron is 0.78 times the speed of light. Strictly speaking, the elastic scattering of such electrons should be solved with the Dirac equation. However, it has been shown by Rujiwara, Hashimoto and more recently by Rother and Scheerschmidt that a scalar-ralativistic-corrected Schrodinger equation is accurate enough for 300-keV electrons (Fujiwara, 1961; Hashimoto, 1964; Rother & Scheerschmidt, 2009). In addition, for high-energy electrons, the exchange and correlation with the electrons in the solids can be neglected (Mott & Massey, 1965; Wu & Ohmura, 1962). Therefore, the elastic scattering of an electron can be described by a Schrodinger equation for the atomic potential (the Coulomb potential of the nuclei as screened by the innershell and valence electrons) (Cowley & Moodie, 1957; Fujiwara, 1961; Gratias & Portier, 1983; Humphreys, 1979) :

$$\left[-\frac{\hbar^2}{2m}\nabla^2 - eV(\mathbf{r})\right]\Psi(\mathbf{r}) = E\Psi(\mathbf{r}) \tag{1}$$

where $m = \gamma m_0$ is the relativistic mass, $-e$ is the charge of an electron, $V$ is the screened atomic potential inside the sample, and $E$ is the incident kinetic energy of the electron.

To solve forward propagation of high-energy electrons in *real space* governed by Eq. (1), we need to make two approximations. The first one is to assume the wave function $\Psi(x, y, z)$ can be written as a product of two factors–one fast varying term along $z$, $\exp(2\pi i z/\lambda)$, and one slow-varying envelope term $\psi(x, y, z)$ (Cowley & Moodie, 1957; Kirkland, 2010; Van Dyck & Coene, 1984)

$$\Psi(x, y, z) = \psi(x, y, z)\exp(2\pi i z/\lambda) \tag{2}$$

where $\lambda$ is the relativistic wavelength of the high-energy electron ($\lambda = \frac{h}{\sqrt{2\gamma m_0 E}}$). The second approximation is the forward scattering approximation (Kirkland, 2010; Van Dyck & Coene, 1984):

$$\left|\frac{\partial^2 \psi}{\partial z^2}\right| \ll \left|\frac{1}{\lambda}\frac{\partial \psi}{\partial z}\right| \tag{3}$$

The two approximations are valid because the electrons are traveling at very high speed and the deflection by a single atomic potential is relatively weak. Based on the two approximations, if we substitute Eq. (2) into Eq. (1) and drop the $\frac{\partial^2 \psi}{\partial z^2}$ term, we have:

$$i\hbar \frac{h}{m\lambda}\frac{\partial \psi(x,y,z)}{\partial z} = \left[-\frac{\hbar^2}{2m}\nabla^2_{xy} - eV(x, y, z)\right]\psi(x, y, z) \tag{4}$$

The propagation of high-energy electrons governed by Eq. (4) can be approximately solved by the multislice method (Chen & Van Dyck, 1997; Gratias & Portier, 1983; Kirkland, et al., 1987; Van Dyck & Coene, 1984).

Apart from the fact a column of atoms is $z$ dependent, a few bound states are supported on the column (Anstis, et al., 2003; Nellist & Pennycook, 1999; Pennycook & Jesson, 1990). It is a valid approximation to expand the atomic column into a Fourier series and keep only the z-independent zero-order term, i.e. the average potential of the atomic column $V(x,y) = \frac{v_{\Delta z}(x,y)}{d}$. This is the same as the approximation of keeping only the zero-order Laue zone in the Bloch-wave simulation (Berry & Ozoriode.Am, 1973). The scattering to the first, second and higher Laue zones can be added later using time-dependent perturbations(Gratias & Portier, 1983). The $z$-independent potential converts Eq. (4) into a 2-D time-independent problem(Berry & Ozoriode.Am, 1973):

$$\left[-\frac{\hbar^2}{2m}\nabla^2_{xy} - eV(x,y)\right]\psi_i(x,y) = E_i\psi_i(x,y) \qquad (5)$$

If the incident wave function at the entrance surface is $\Psi(x,y)$, then the depth evolution of the wave function is

$$\psi(x,y,z) = \sum_i C_i \psi_i(x,y)\exp(-\frac{i}{\hbar}\frac{m\lambda}{h}E_i z) \qquad (6)$$

$$C_i = \int \Psi(x,y)\psi_i(x,y)dxdy \qquad (7)$$

In Eq. (7), $C_i$ is the coupling coefficient of the incident probe ($\Psi(x,y)$) to a particular columnar orbital $\psi_i(x,y)$. When the incident probe $\Psi(x,y)$ is cylindrically symmetric, as is in a regular STEM, and positioned on the column, the coupling coefficient is zero for any columnar states with non-zero angular momentum such as $2p$. Now, we can transfer our knowledge of solving bound orbitals of 3-D atoms (Cowan, 1981; Koonin & Meredith, 1998) to the problem of solving a 2-D projected screened "atom".

## The Radial Equation

First, for computational convenience, we use the Ry-Bohr atomic units:

$$Ry = \frac{\hbar^2}{2m_0 a_0^2} = 13.607 \text{ eV}$$

$$a_0 = \frac{\hbar^2}{m_0 e^2} = 0.529 \text{ Å}$$

Then Eq. (5) for a single column in cylindrical coordinates reads:

$$\nabla^2 \psi(r,\theta) + \gamma[E + V(r)]\psi(r,\theta) = 0 \tag{8}$$

Where $\gamma$ is the relativistic gamma. $V(r)$ is the potential energy of the electron in Ry and it is positive at finite distances. Because for a single column $V$ is cylindrically symmetric, we can perform separation of variables and has the following form:

$$\begin{aligned}\psi_{nl}(r,\theta) &= c_l u_{nl}(r) \exp(il\theta) \\ &= c_l \frac{R_{nl}(r)}{r^{1/2}} \exp(il\theta)\end{aligned} \tag{9}$$

where $l$ is the angular quantum number or the topological charge and it can be any integer from $-\infty$ to $+\infty$. The radial wave equation for $R_{nl}(r)$ is:

$$\frac{d^2 R_{nl}(r)}{dr^2} + \left[\gamma E + \gamma V(r) - \frac{l^2 - \frac{1}{4}}{r^2}\right] R_{nl}(r) = 0 \tag{10}$$

Except for the relativistic correction, (Berry & Ozoriode.Am, 1973) showed the same equation.

## 2p Columnar Orbital

For *p*-type columnar orbitals that carry topological charge one ($l = 1$), Eq. 10 becomes:

$$\frac{d^2 R_{n1}(r)}{dr^2} + \left[\gamma E + \gamma V(r) - \frac{3/4}{r^2}\right] R_{n1}(r) = 0 \tag{11}$$

It is easy to obtain the asymptotic forms of $R_{n1}(r)$:

$$R_{n1}(r) = c \exp\left[-\sqrt{\gamma |E_{n1}|}\, r\right] \quad as\ r \to \infty$$
$$R_{n1}(r) = c r^{3/2} \quad as\ r \to 0 \tag{11}$$

As $u_{nl}(r) = cr \to 0\ as\ r \to 0$, the amplitude of the $p$ columnar orbitals is zero in the center of the orbital. Therefore, they are on-column vortex beams hosted by the crystal with topological charge one.

As we know the initial conditions from the asymptotic forms, the radial function $R_{n1}$ can be calculated by numerical integration. A very primitive method in solving such eigenenergy problem is the shooting method: the eigenstates can be identified when the radial function does not blow up at large distances. An improved method is to both integrate forward from $r = 0$ and integrate backward from $r = \infty$. The eigenstate can be identified when the first derivative at the connection point is continuous. The numerical integration of this 2$^{nd}$ order differential equation can be performed by a Runge-Kutta method. However, similar to solving 3-D atomic orbitals, the more efficient Numerov method can be applied (Cowan, 1981; Koonin & Meredith, 1998).

Figure 5 shows the numerical solutions for two typical columns that can be found in $3d$ transition metals and in perovskite materials. Take cobalt for example, the $2p$ columnar state is plotted in Figure 5b. It is clearly a vortex beam with topological charge +1. The full width at half maximum is only 1.22 Å. In the lateral dimension, the nearest column is positioned 2.2 Å away. This gives an overlap integral smaller than 1% which means the $2p$ columnar orbital can be considered isolated without interference from the neighbors. If you compare the $2p$ orbital on the Co column with the 300 kV vortex beam created by an aberration-corrected electron microscopy, the size matches well. This means that for the on-column 2*p* orbital of a given material and a given zone axis, the convergence angle can be optimized such that the size of the vortex beam matches that of the 2*p* columnar orbital to maximize excitation. Figure 6 gives such guidelines for two typical elements in 3*d* and rare earth. One can look up the FWHM of the 2*p* orbitals in this plot and use the equation in Figure 2 to calculate the optimal convergence angle for maximal coupling.

## On-Column Channeling of Vortex Beams

Figure 7 shows the propagation of the central beam with the vortex side bands with a column of cobalt placed in the core of the left vortex. In Figures 7a and 7b, we can see the left vortex channels along the column whereas the central and right beams diverge as they propagate. The 2*p* state propagates along the column and the topological charge is preserved even at a thickness of 50 nm as shown in the phase map.

2*p* columnar orbitals can also be excited when a round beam is placed off the column. However, compared to the excitation of 1*s*, only a minimal fraction is couple to the *p*-type bound states due to the phase symmetry. We want to note that that such strong and selective excitation of 2*p* columnar orbitals is only possible with an incident beam that possesses the same orbital angular momentum.

# DISCUSSIONS

Even though we focused the discussion on 2*p* columnar states, the vortex beam of topological charge one can excites higher *p* states. However, the isolated column model for higher *p* states does not hold as well as it does for 2p orbitals because they are more delocalized. Even for 2p columnar orbital, the isolated model can break down when the crystal is oriented to low-symmetry zone axis or light-element crystals are used. To apply the single column model, one needs to calculate the FWHM of the $p$ orbital and make sure that it is smaller than the smallest column-to-column spacing in the lateral dimension. Nevertheless, as we have shown in Figure 5, in two typical materials made of 3d and rare earth elements oriented along high-symmetry zone axis, the high atomic density along the column attracts the 2p columnar orbital tight enough such that the overlap with the neighbor orbitals is ignorable. In general, high-symmetry zone axis is preferred as the lateral column-to-column spacing is high and the interatomic distance along the column is low.

As shown by McMorran *et al*, electron vortex beams with high quanta of orbital angular momentum can be intensively generated using holograms of high-quanta helical phase fronts (McMorran, et al., 2011b). This gives possibility to selectively excite columnar orbitals with absolute topological charge larger than one. However, one has to check if such orbitals are supported on the columns and if the isolated column approximation is valid.

In this paper, we assume the optics is aberration free. This assumption is approximately correct in an aberration-corrected instrument with moderately chosen numerical aperture. However, in an uncorrected microscope, a modeling with the third order spherical aberration is needed. In addition to the geometric aberrations, chromatic aberration can cause defocus blur to the beams. This effect can be important in instruments with large $C_c$ and large energy spread(Intaraprasonk, et al., 2008).

# CONCLUSION

In this paper, we first reviewed the forming of an atomic vortex beam in an aberration-corrected scanning transmission electron microscope. We presented the theoretical scalings for various characteristic sizes, including peak-intensity radius, the full width at half maximum and the Rayleigh radius, of the vortex beam with topological charge one. We also showed that the vortex side band diverges as well as the central beam as they propagate away from the optimal focal plane. The helical phase front can be destroyed when the central beam interferes with the side vortex bands. We developed a simple formula to calculate the least number of fringes needed in the pitchfork aperture for the distance that the vortex beam can travel without strong interferences from the central beam. Then we reviewed the theory for calculating the bound states on single atomic columns. We gave numerical solutions to the $2p$ columnar orbitals and we tabulated the FWHM of the $2p$ states of two typical elements, one $3d$ transition and one rare earth, of various incident beam energies and various interatomic distances along the column. The tabulated numbers can help to estimate the optimal convergence angle for maximal coupling to $2p$. Finally, we showed the vortex beam can channels on atomic columns and preserve the topological charge of the incident beam by coupling to the $2p$ columnar orbital with the same angular momentum. This shows that the vortex nature of a vortex beam can be extended and the beam divergence can be counteracted by an atomic column.

# ACKNOWLEDGEMENTS

This work was supported by Materials Sciences Division, Lawrence Berkeley National Lab. HLX thanks his thesis advisor David A. Muller for giving the project and countless advice on calculating columnar orbitals for electron channeling during his Ph.D. study. HLX thanks Judy J. Cha and Earl J. Kirkland for initial code and notes for solving the radial Schrödinger equation. HLX also thanks Robert Hovden for helping debug the program and developing the concept of columnar orbitals.

# FIGURE CAPTIONS

Figure 1. The forming of atomic-size vortex beams with topological charge one. (a) The optical schematic. (b) The amplitude and the phase of the central and the vortex side bands at the optimal defocus (300 keV, $\alpha_{max} = 20\ mrad$). The pitchfork hologram aperture was generated by threshholding the interference pattern generated by coherent superposition of a helical phase front and a plane wave. We chose the same number of fringes as used in ref (Verbeeck, et al., 2010).

Figure 2. The scaling of the sizes of a vortex beam as a function of wavelength and convergence semi-angle. (a) The radial amplitude of a vortex beam with topological charge one (300 keV, $\alpha_{max} = 20\ mrad$). (b) Various characteristic dimensions as a function of the convergence semi-angle.

Figure 3. The free propagation of the vortex beams (300 keV, $\alpha_{max} = 20\ mrad$) in vacuum. The amplitude and the phase of the wave function at (a) 0 nm, (b) 5 nm, (c) 10 nm, (d) 15 nm and (e) 20 nm away from the optimal focal plane. (f) The schematics of the free propagation of vortex beams.

Figure 4. The least number of fringes required to minimize interference from the central beam as the vortex beam propagates. The plots show the calculations for (a) 60 keV, (b) 100 keV, (c) 200 keV, and (d) 300 keV.

Figure 5. The 2*p* orbital with topological charge +1 of the (a-b) cobalt column with 2.5074-Angstrom spacing (columns down hexagonal cobalt [11-20]) and the (c-d) dysprosium column with 4-Angstrom spacing (pseudocubic perovskite [001] axis).

Figure 6. The FWHM of the 2*p* orbital hosted on (a) cobalt and (b) dysprosium columns at a function of the interatomic distance along the vertical column at a few typical incident beam energies. This serves as a guideline for the choice of the convergence angle for the vortex beam to maximize coupling to 2*p*.

Figure 7. The multislice simulation of the propagation of the central beam and the vortex side bands with a cobalt column (2.5074-Angstrom spacing, the column down hexagonal cobalt [11-20]) positioned in the center of the left vortex beam. The intensity and the phase of the wave function as it propagates (a) 4.25 nm, (b) 15.25 nm, (c) 37.25 nm, and (d) 50.25 nm along the column. (e) The schematics of the propagation the vortex beams. The simulation was carried out by a custom-written matlab script.

# Figures

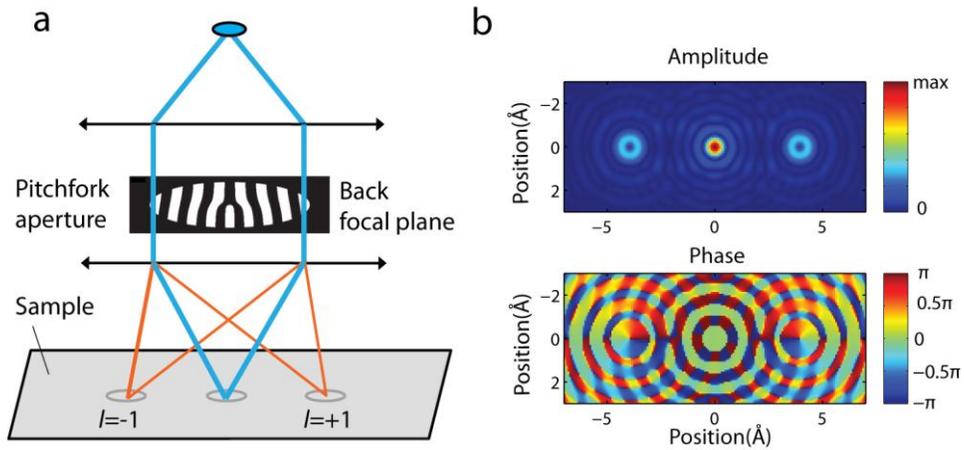

Figure 1. The forming of atomic-size vortex beams with topological charge one. (a) The optical schematic. (b) The amplitude and the phase of the central and the vortex side bands at the optimal defocus (300 keV, $\alpha_{max} = 20\ mrad$). The pitchfork hologram aperture was generated by threshholding the interference pattern generated by coherent superposition of a helical phase front and a plane wave. We chose the same number of fringes as used in ref (Verbeeck, et al., 2010).

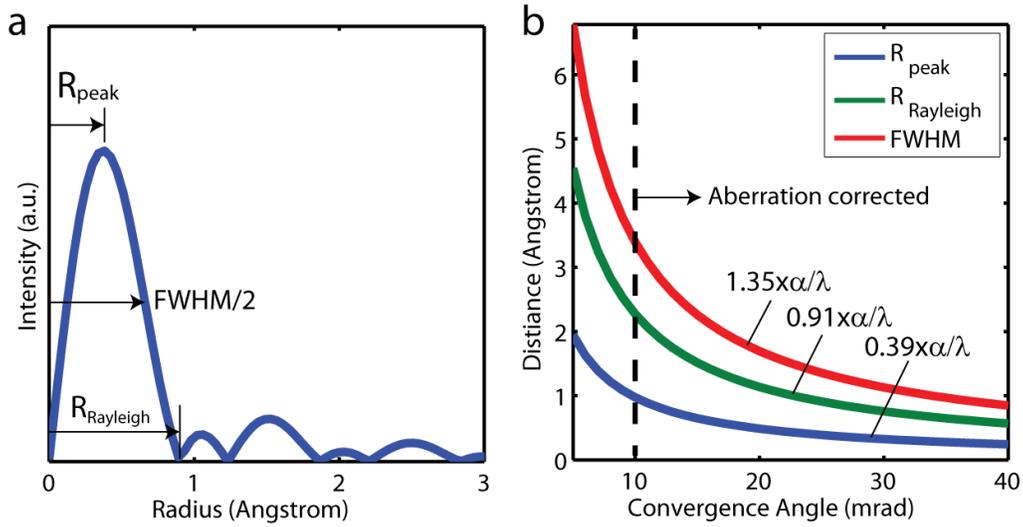

Figure 2. The scaling of the sizes of a vortex beam as a function of wavelength and convergence semi-angle. (a) The radial amplitude of a vortex beam with topological charge one (300 keV, $\alpha_{max} = 20\ mrad$). (b) Various characteristic dimensions as a function of the convergence semi-angle.

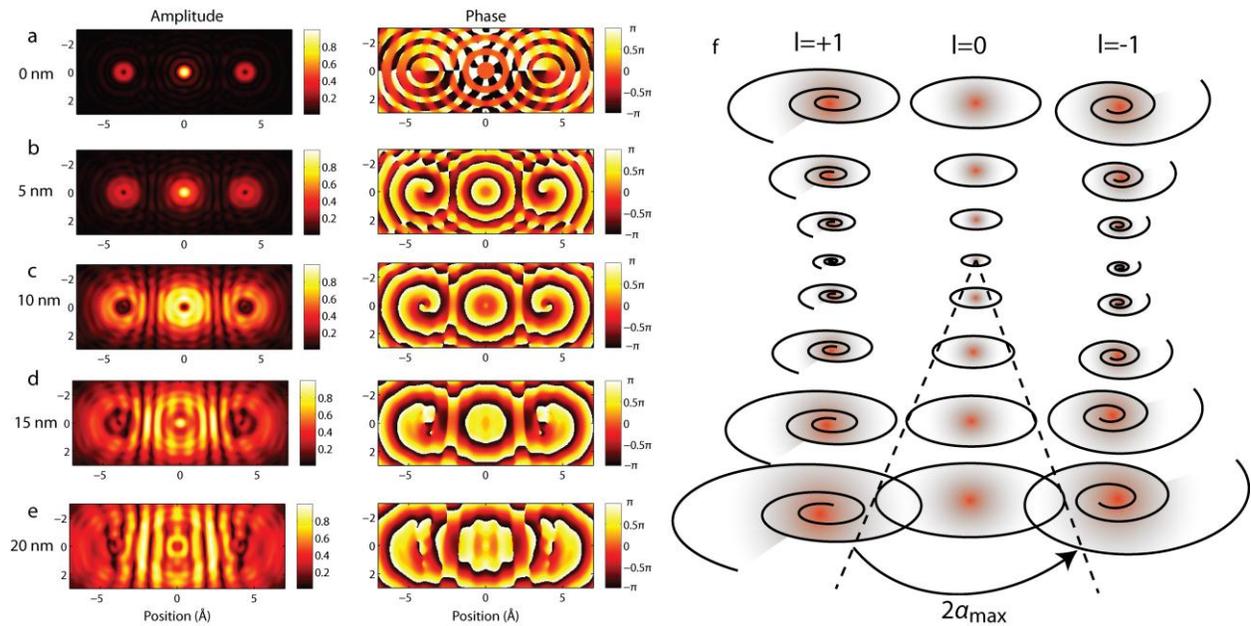

Figure 3. The free propagation of the vortex beams (300 keV, $\alpha_{max} = 20\ mrad$) in vacuum. The amplitude and the phase of the wave function at (a) 0 nm, (b) 5 nm, (c) 10 nm, (d) 15 nm

and (e) 20 nm away from the optimal focal plane. (f) The schematics of the free propagation of vortex beams.

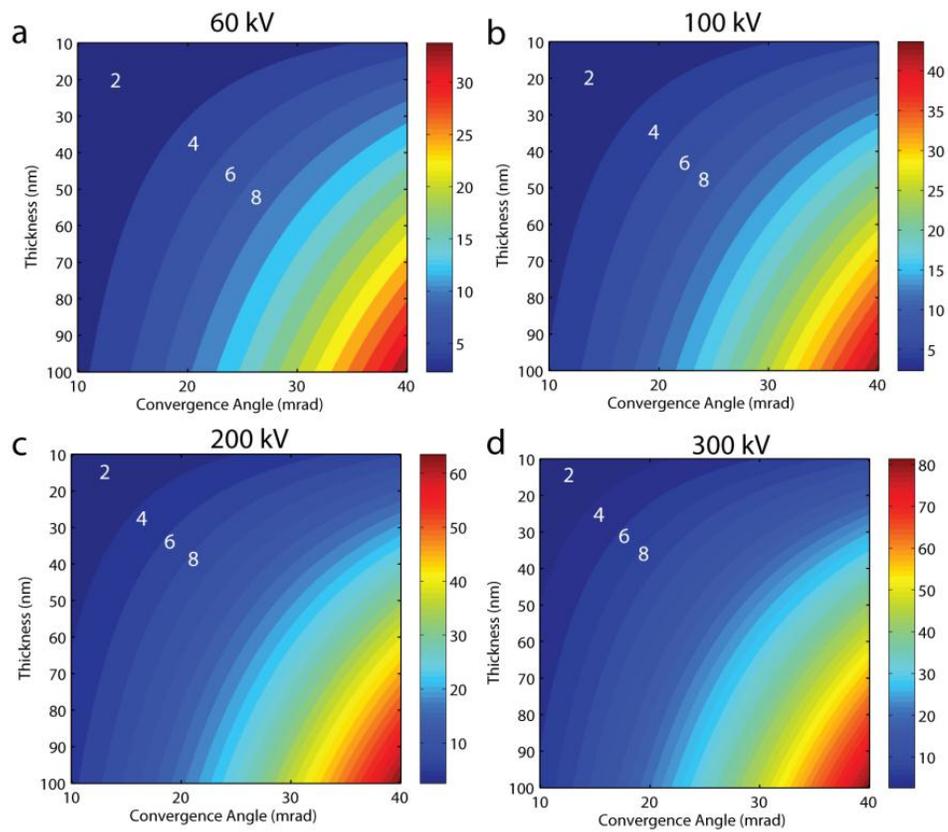

Figure 4. The least number of fringes required to minimize interference from the central beam as the vortex beam propagates. The plots show the calculations for (a) 60 keV, (b) 100 keV, (c) 200 keV, and (d) 300 keV.

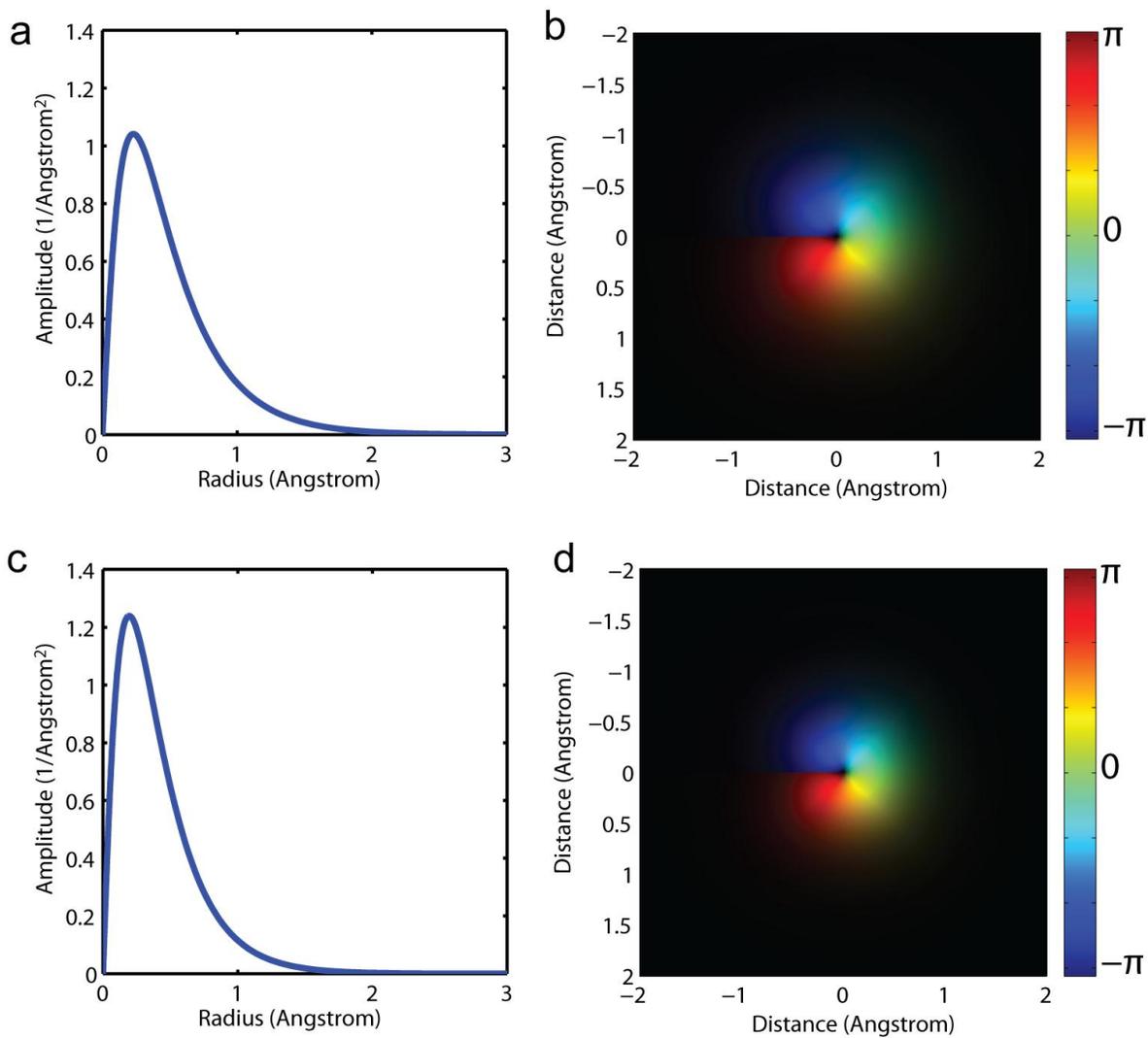

Figure 5. The 2*p* orbital with topological charge +1 of the (a-b) cobalt column with 2.5074-Angstrom spacing (columns down hexagonal cobalt [11-20]) and the (c-d) dysprosium column with 4-Angstrom spacing (pseudocubic perovskite [001] axis).

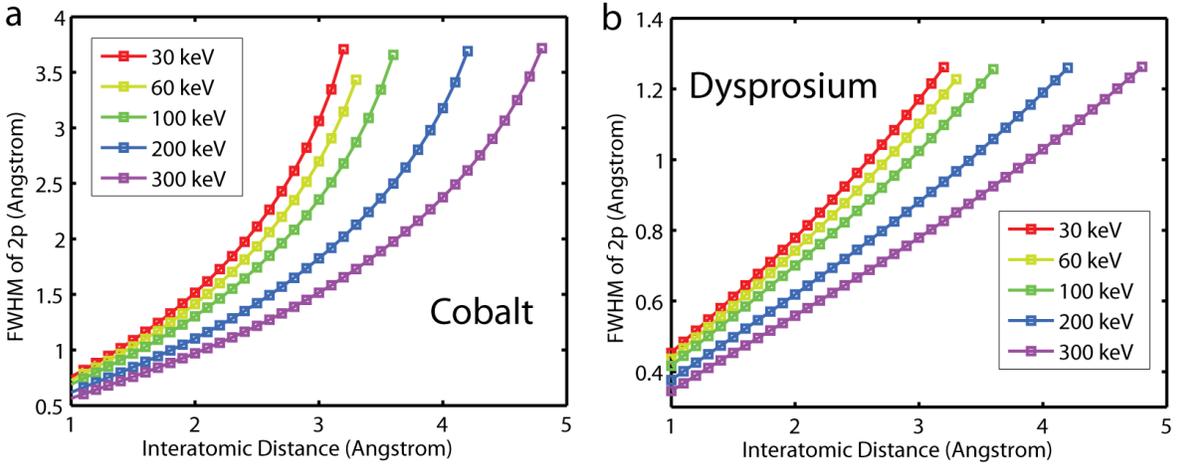

Figure 6. The FWHM of the 2*p* orbital hosted on (a) cobalt and (b) dysprosium columns at a function of the interatomic distance along the vertical column at a few typical incident beam energies. This serves as a guideline for the choice of the convergence angle for the vortex beam to maximize coupling to 2*p*.

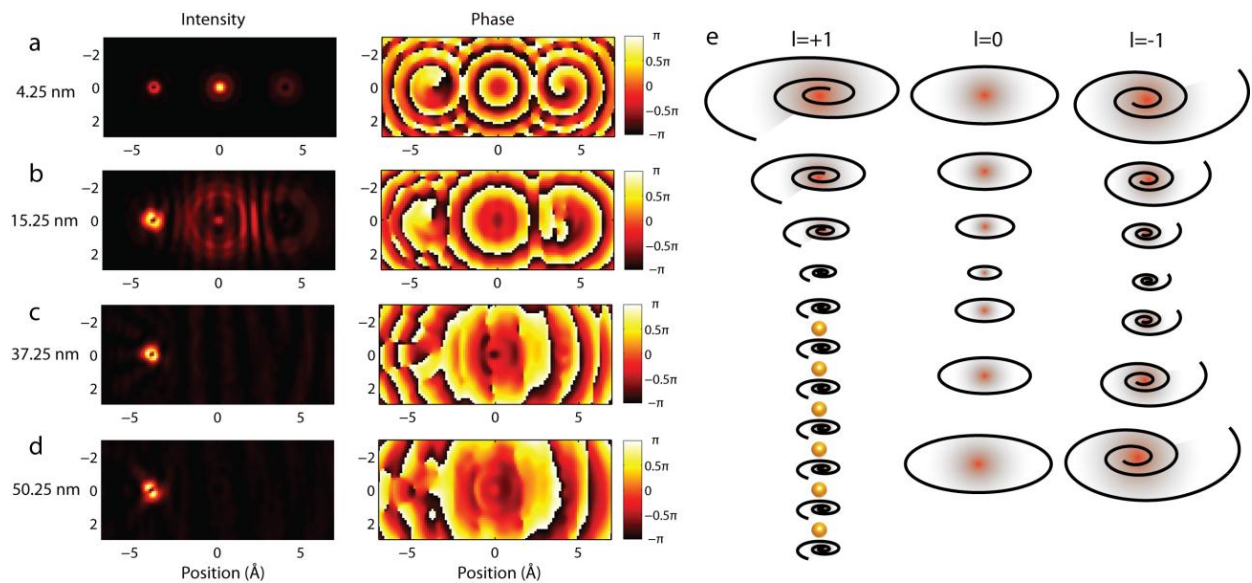

Figure 7. The multislice simulation of the propagation of the central beam and the vortex side bands with a cobalt column (2.5074-Angstrom spacing, the column down hexagonal cobalt [11-20]) positioned in the center of the left vortex beam. The intensity and the phase of the wave function as it propagates (a) 4.25 nm, (b) 15.25 nm, (c) 37.25 nm, and (d) 50.25 nm along the column. (e) The schematics of the propagation the vortex beams. The simulation was carried out by a custom-written matlab script.